\documentclass[twocolumn,preprintnumbers,amssymb,amsmath,superscriptaddress,letterpaper]{revtex4}[10pt]

\usepackage{graphicx}
\usepackage{dcolumn}
\usepackage{bm}
\usepackage{axodraw}
\usepackage{pstricks}
\usepackage{color}
\usepackage{epsfig}

\advance\textheight	 0.4 in
\advance\textwidth 0.2in

\newcommand{\be}{\begin{equation}}
\newcommand{\ee}{\end{equation}}
\newcommand{\bea}{\begin{eqnarray}}
\newcommand{\eea}{\end{eqnarray}}

\newcommand{\gsim}{ \mathop{}_{\textstyle \sim}^{\textstyle >} }
\newcommand{\lsim}{ \mathop{}_{\textstyle \sim}^{\textstyle <}}

\newcommand{\gev}{{\rm ~GeV}}

\def\draft{
}

\begin{document}
\draft

\title{The PAMELA Positron Excess from Annihilations into a Light Boson}

\author{Ilias Cholis}
\affiliation{Center for Cosmology and Particle Physics, Department of Physics, New York University, 
New York, NY 10003}

\author{Douglas P. Finkbeiner}
\affiliation{Harvard-Smithsonian Center for Astrophysics, 60 Garden St., Cambridge, MA 02138}

\author{Lisa Goodenough}
\affiliation{Center for Cosmology and Particle Physics, Department of Physics, New York University, 
New York, NY 10003}

\author{Neal Weiner}
\affiliation{Center for Cosmology and Particle Physics, Department of Physics, New York University, 
New York, NY 10003}

\date{\today}

\begin{abstract}
Recently published results from the PAMELA experiment have shown conclusive evidence for an excess of positrons at high ($\sim 10 - 100 \, \gev$) energies, confirming earlier indications from HEAT and AMS-01. Such a signal is generally expected from dark matter annihilations. However, the hard positron spectrum and large amplitude are difficult to achieve in most conventional WIMP models. The absence of any associated excess in anti-protons is highly constraining on models with hadronic annihilation modes. We revisit an earlier proposal, wherein the dark matter annihilates into a new light $(\lsim \gev)$ boson $\phi$, which is kinematically constrained to go to hard leptonic states, without anti-protons or $\pi^0$'s. We find this provides a very good fit to the data. The light boson naturally provides a mechanism by which large cross sections can be achieved through the Sommerfeld enhancement, as was recently proposed. Depending on the mass of the WIMP, the rise may continue above 300 GeV, the extent of PAMELA's ability to discriminate between electrons and positrons.
   
\end{abstract}
\maketitle

\section{Introduction}
\label{sec:intro}

Thermal WIMPs are attractive as an explanation of the cold dark matter
of the universe. The thermal relic freeze-out argument provides both
an order of magnitude estimate of the relic abundance of the dark
matter, and a mechanism for observational signatures. The annihilation of the WIMPs into standard model states naturally produces a new source of cosmic rays, to which a variety of experiments are sensitive.

Particularly interesting searches are those for anti-matter cosmic rays. Anti-matter is expected to be produced from standard model interactions of high energy protons with the interstellar medium. Generically, the spectrum of anti-matter is expected to fall relative to matter as one moves to high energies. A sharp upturn in the ratio of the two would be a strong indication of a new primary source of cosmic rays \cite{Serpico:2008te}.

The recently released positron data \cite{Adriani:2008zr} from the PAMELA experiment \cite{Boezio:2008mp} provides an exceptionally strong case for just such a source of high energy positrons.

Indeed, anticipation of the PAMELA data release has generated a significant amount of excitement \cite{cirelli,bergstrom,Cirelli:2008pk,Cholis:2008hb,Huh:2008vj,Barger:2008su,Chen:2008yi,Chen:2008dh,Fairbairn:2008fb,Hooper:2008kg,Yuksel:2008rf}. Dark matter annihilations have long been considered as a possible source of  a positron excess \cite{Turner:1989kg,Kamionkowski:1990ty,Jungman:1995df, darkpositron0,Baltz:2001ir,Kane:2002nm,Cheng:2002ej,Hooper:2004xn,Hooper:2004bq,pierce,Nagai:2008se}. Alternative sources, such as pulsars \cite{pulsars,pulsars2, pulsars3,Hooper:2008kg,Yuksel:2008rf} provide an alternative explanation of these positrons, and so it behooves us to consider the implications of this signal for any putative WIMP candidate.

The most notable (and convincing) feature of the signal is the incredibly sharp rise, with the positron fraction $e^+/(e^++e^-)$ rising by a factor of 2.5 from 10.77 to 82.55 GeV. Indeed, in the present data, there is no indication of a flattening. Such a rapid rise, interpreted as a WIMP annihilation signal is difficult to understand in terms of most conventional supersymmetric models. Such hard spectra, interpreted as annihilations directly to standard model states, seem to favor annihilations straight to leptons \cite{Cholis:2008hb,Cirelli:2008pk}. Direct annihilations to light leptons, however, are helicity suppressed for Majorana fermions such as the neutralino \footnote{Annihilations to $e^+e^- \gamma$ need not be similarly suppressed, although they are still typically small \cite{bergstrom}.}. Moreover, the PAMELA mission sees no excess in anti-protons \cite{Adriani:2008zq}, while annihilations through hadronic channels generally give copious anti-protons above backgrounds \cite{darkantiproton1,darkantiproton2}, making these annihilations severely constrained, if not excluded, as an explanation of such an excess \cite{Cirelli:2008pk}.

In this note, we will reexamine a previously proposed \cite{Cholis:2008vb,ArkaniHamed:2008qn} annihilation channel for high energy positrons. Namely, we will consider annihilations into a new, light $(m_\phi \lsim \gev)$ boson, which could be a scalar or vector. Because the boson is so light, it is kinematically constrained to decay to dominantly leptonic modes, producing a hard spectrum of positrons, without any additional anti-protons. We will see that the previously predicted spectra give an excellent fit to the data, and may continue to extend up to the maximum range of the PAMELA reach, depending on the mass of the WIMP.
\vskip 0.15in
\section{Annihilations to a New Light Boson}
\begin{figure}
\begin{center}
\scalebox{0.6}{
\includegraphics[width=3.1in,angle=0]{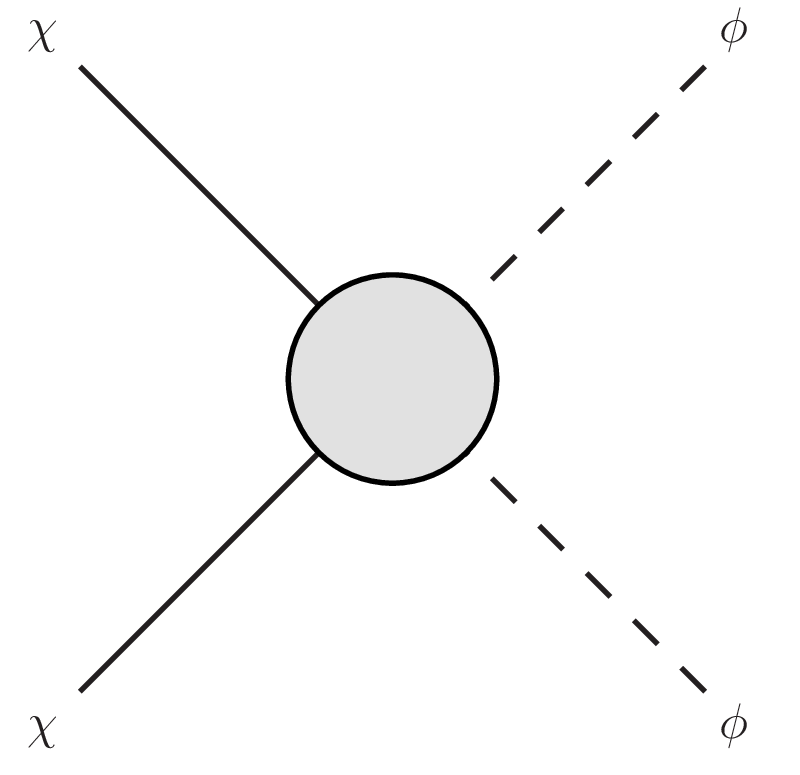}
}
\end{center}
\caption{Dark matter annihilation mode, $\chi \chi \rightarrow \phi\phi$. Here $\phi$ can be a scalar or a vector. The grey circle includes all diagrams, including possible non-perturbative contributions. The subsequent decays of $\phi$, for instance to $\phi \rightarrow e^+e^-$ or $\phi \rightarrow \mu^+ \mu^-$ produce the high energy positrons observed at PAMELA.}
\label{fig:annihilation}
\end{figure}

In \cite{Cholis:2008vb}, Cholis, Goodenough and Weiner studied the possible implications of the XDM model of Finkbeiner and Weiner \cite{Finkbeiner:2007kk}. Because a new light force carrier is required by the theory, it provides a new annihilation channel, namely $\chi \chi \rightarrow \phi\phi$ (Fig. \ref{fig:annihilation}). Subsequent decays of the $\phi$ provide the excess positrons observed at PAMELA, HEAT \cite{Barwick:1997ig}, AMS-01 \cite{AMS} and other experiments.

Because the $\phi$ is light, decays are kinematically prevented from producing anti-protons, but instead produce highly boosted leptons. More recently, it has been argued that a light boson naturally provides a significant enhancement at the low velocities found in the halo from the Sommerfeld enhancement \cite{ArkaniHamed:2008qn,ArkaniHamed:2008qp}\footnote{The importance of the Sommerfeld enhancement for WIMP annihilation was first discussed by \cite{Hisano:2003ec,Hisano:2004ds} in the context of heavy WIMPs, where the force carrier was the standard model W and Z bosons. More recently, it was examined in the context of ``minimal dark matter'' as well \cite{Cirelli:2007xd,Cirelli:2008id,cirelli}, and annihilations through the Higgs portal \cite{MarchRussell:2008yu}.}, or recombination into a bound WIMPonium state \cite{Pospelov:2008jd,MarchRussell:2008tu}.  Indeed, precisely this sort of light force carrier was a central element in a recently proposed unified dark matter explanation \cite{ArkaniHamed:2008qn} of INTEGRAL, PAMELA, ATIC, DAMA, and the WMAP ``Haze''.

We should note that the idea that dark matter may couple to some additional boson is hardly radical. The possibility of a $U(1)$ associated with the dark matter was considered in \cite{Holdom:1985ag}, and also arises in ``mirror'' dark matter \cite{Foot:1995pa}. New forces are necessarily invoked for MeV-scale dark matter \cite{Boehm:2003hm,Boehm:2003bt}, as well as ``exciting'' dark matter \cite{Finkbeiner:2007kk}, whose setup we are presently considering. ``WIMPless'' models \cite{Feng:2008ya} have additional interactions and annihilations, as do the general class of ``secluded'' dark matter models \cite{Pospelov:2007mp} (which includes XDM as an example). The only potentially peculiar feature one may wonder about here is that we invoke a light $(m_\phi \lsim {\rm GeV})$ boson, although this scale can arise naturally in supersymmetric theories \cite{ArkaniHamed:2008qp}.

Putting aside model building questions, which have been adequately addressed elsewhere, we focus on the phenomenology of this scenario. Following the annihilation $\chi \chi \rightarrow \phi \phi$, we consider here four cases of $\phi$ decays, specifically: a) $\phi \rightarrow e^+ e^-$, b) $\phi \rightarrow \mu^+ \mu^-$, c) a mixture of 1:1 between electrons and muons, and lastly d)  $\pi^+ \pi^-$. The first is natural for any particle with a mass lighter than $m_\phi < 2 m_\mu$ \footnote{We should note that if the $\phi$ is arbitrarily light, one will encounter significant astrophysical limits \cite{Kamionkowski:2008gj}.}. The decay could arise if $\phi$ is a scalar, mixing with the Higgs, for example \cite{Finkbeiner:2007kk} \footnote{In the case it is a scalar mixing with the Higgs, there is some branching ratio to $\gamma \gamma$, which can rise as high as 14\% just below the $2 m_\mu$ threshold. This can have interesting implications for GLAST, but here changes only the needed boost by a factor of 1/0.86. See \cite{Cholis:2008vb} for a discussion.}, or a vector mixing with the photon \cite{ArkaniHamed:2008qp}. The second (case b) is natural for a scalar mixing with the Higgs in the mass range $2 m_\mu < m_\phi < 2 m_\pi$. 
The third case (1:1) is not precisely realized from photon mixing above $2 m_\mu$, as phase space suppression is a relevant correction until above the $2 m_\pi$ threshold. Rather, it yields something close to a 2:1 ratio of e:$\mu$, which is very similar to the electron only case. A 1:1 ratio could arise in leptophilic models \cite{Fox:2008kb}. Above $2 m_\pi$, the addition of a small pion admixture does not modify the spectrum significantly from 1:1. However, we provide the possibility of a charged pion dominated decay, providing a complete set of low-energy decay modes to consider, in particular in the case that the pion component becomes large. Above $2 m_K$ kaons appear, but hadrons are increasingly inappropriate degrees of freedom, with annihilation to quarks eventually becoming the appropriate description. In such a case, a broad analysis is needed \cite{Meade:2009rb,Mardon:2009rc}.

To calculate the positron spectra and backgrounds, we employ the publicly available GALPROP code \cite{galprop,Galprop1}. We use a Einasto profile \cite{Merritt:2005xc}, with $\alpha = 0.17$. We take a
diffusion zone of $L = 4 {\, \rm kpc}$, a diffusion coefficient at 4MV of  $5.8*10^{28}{\, \rm cm^2/s}$, with an index of
 $0.33$.

\begin{figure*}[t]
\includegraphics[width=3.3in,angle=0]{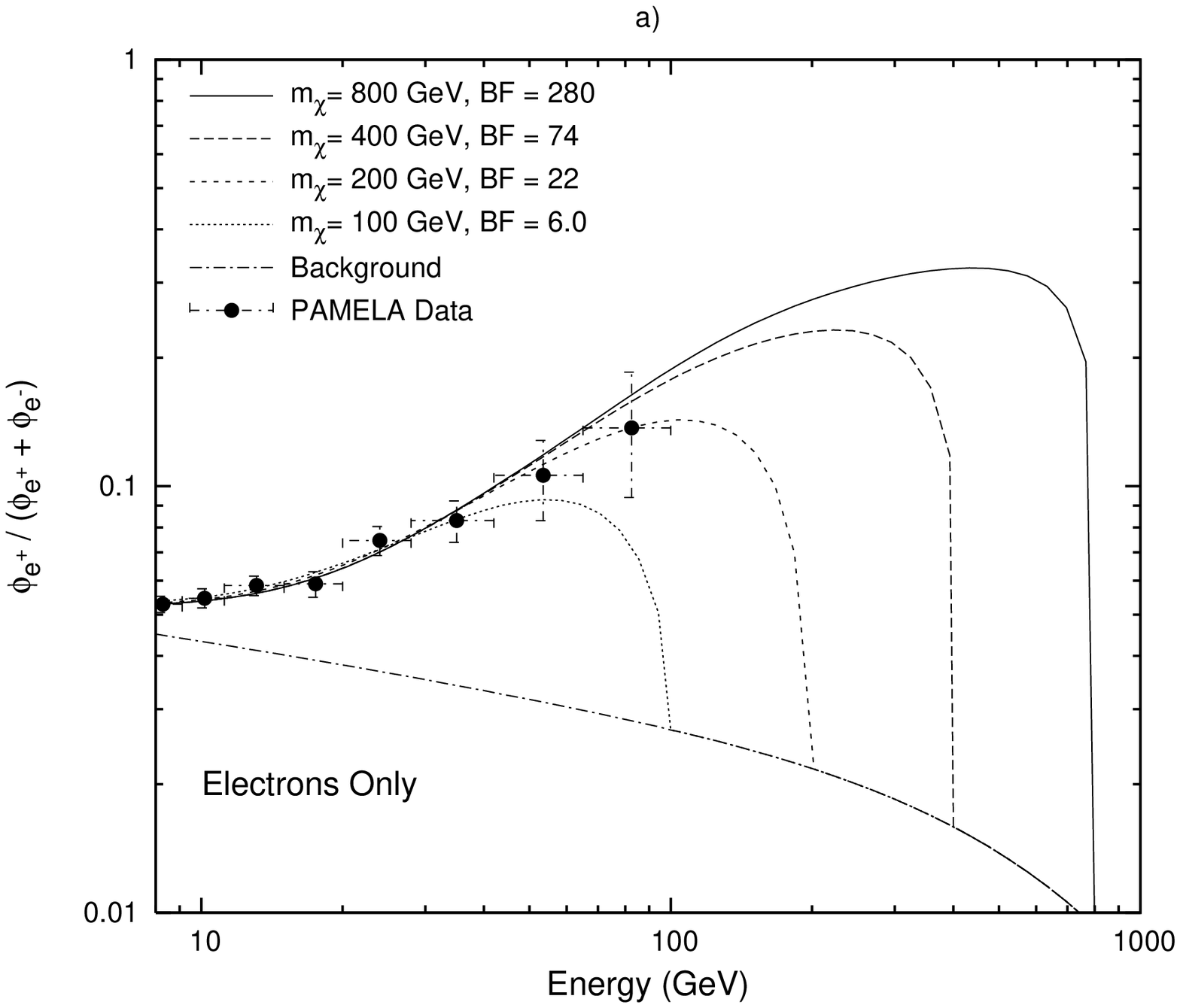}
\includegraphics[width=3.3in,angle=0]{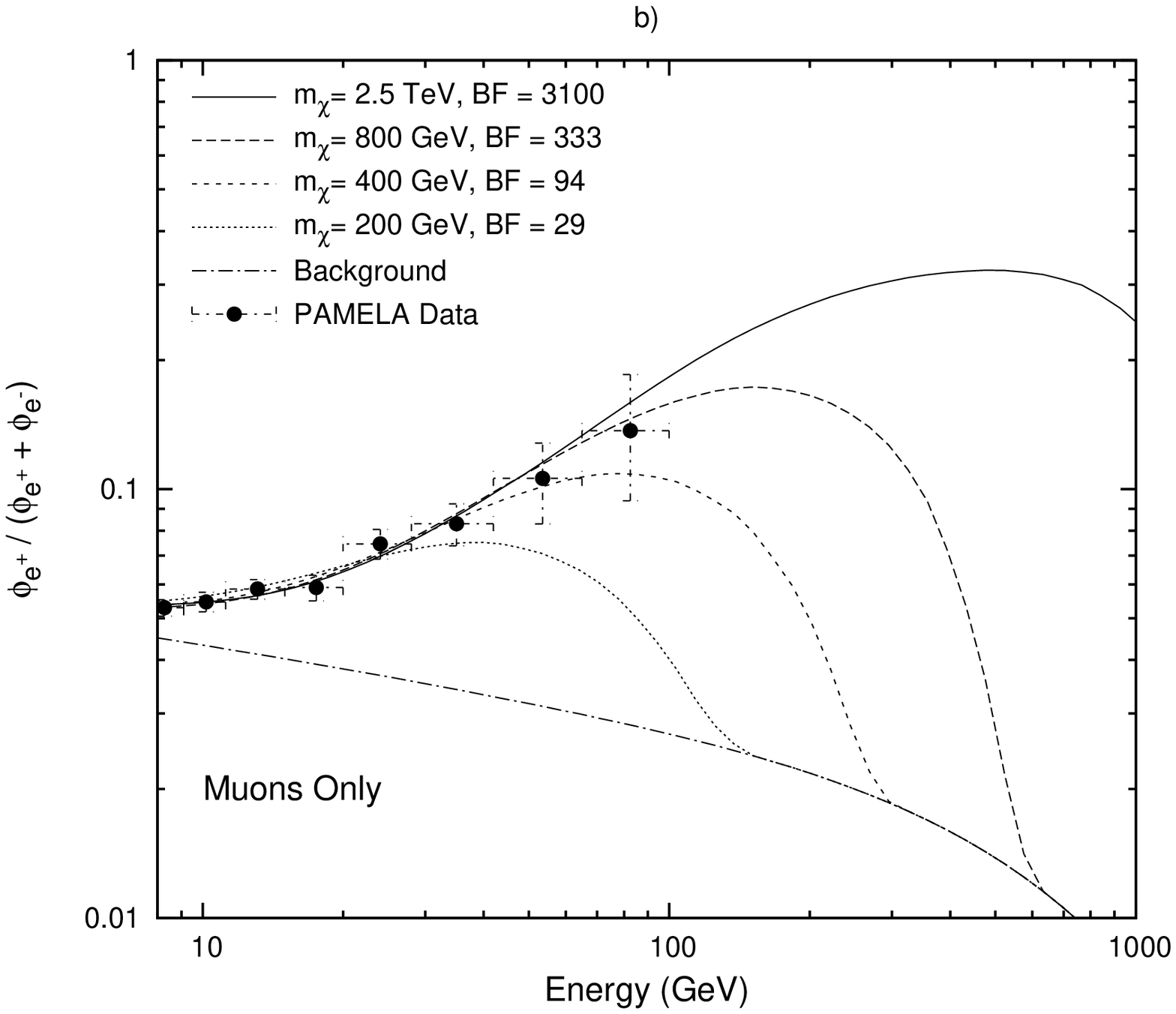}\\
\vskip 0.2cm
\includegraphics[width=3.3in,angle=0]{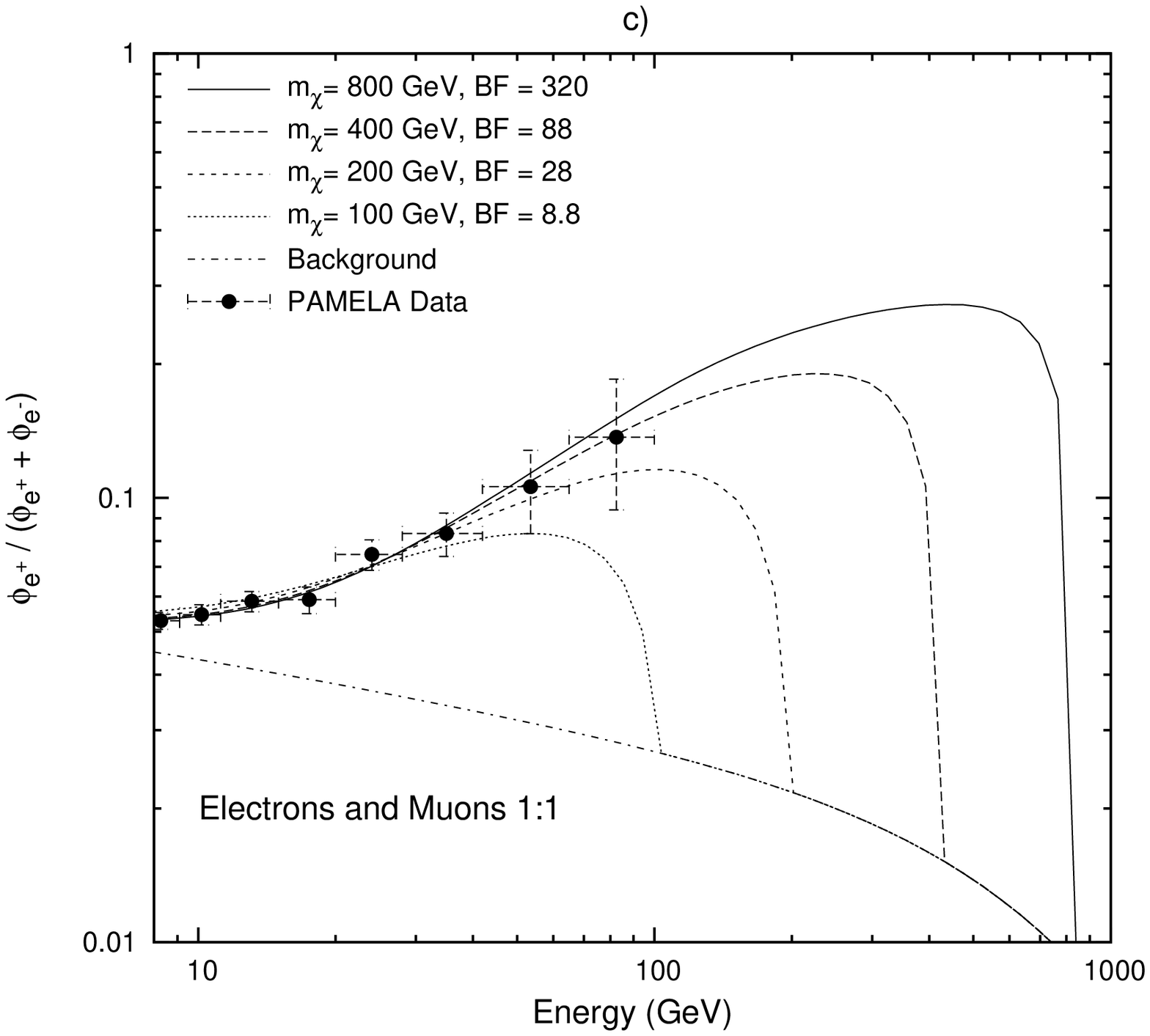}
\includegraphics[width= 3.3in,angle=0]{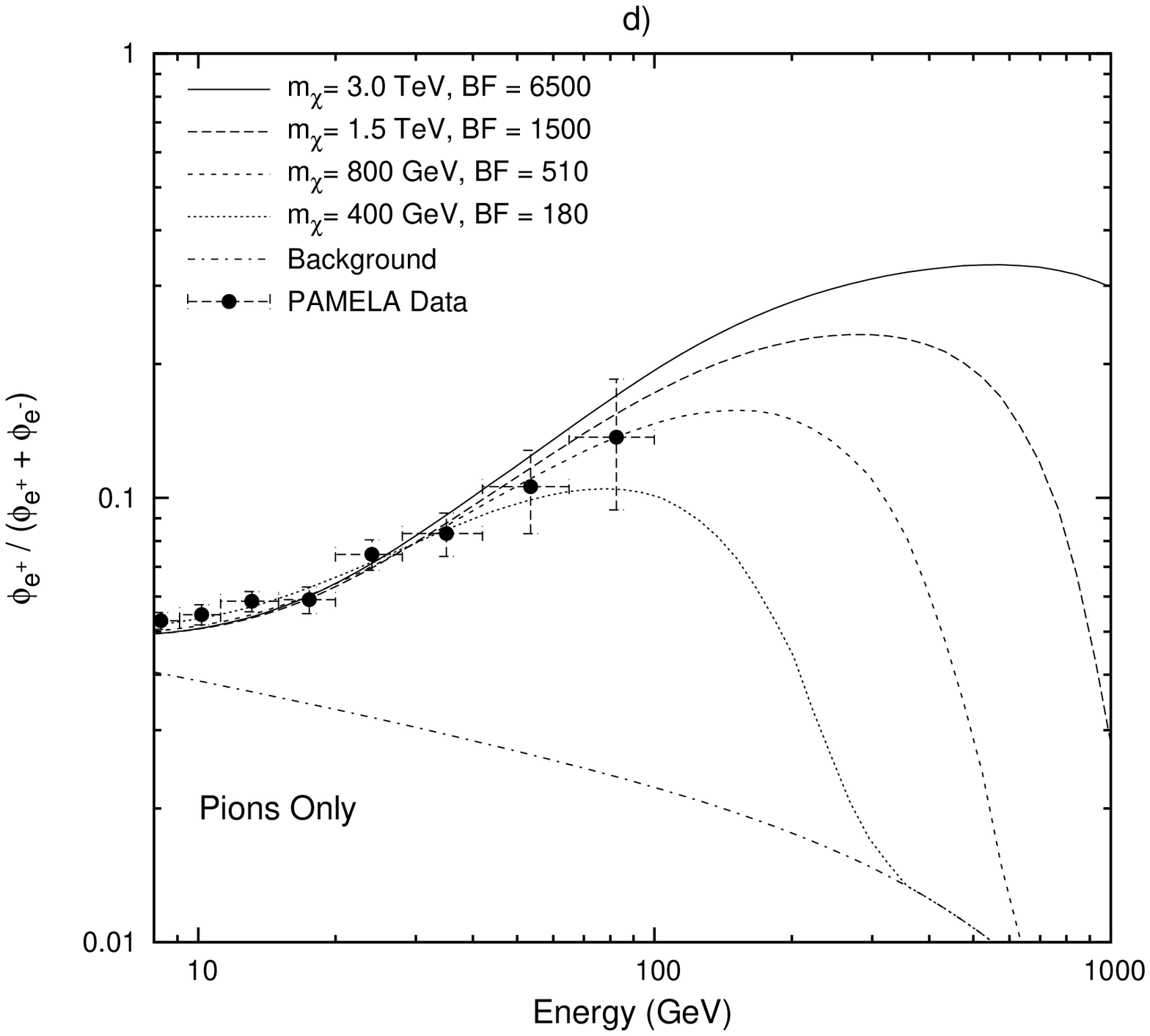}
\caption{The positron fraction as a function of energy for the four annihilation modes considered here: $\chi \chi \rightarrow \phi \phi$, followed by, (a) $\phi \rightarrow e^+e^-$, (b) $\phi \rightarrow \mu^+ \mu^-$, (c) $\phi \rightarrow e^+e^- ,\, \mu^+ \mu^- $ (1:1),  (d) $\phi \rightarrow \pi^+ \pi^-$ . The boost factor is defined relative to a cross section $\langle \sigma v \rangle = 3 \times 10^{-26} {\rm cm^3 s^{-1}}$ and $\rho_0 = 0.3 {\, \rm GeV cm^{-3}}$. Such a boost reasonably can arise from a Sommerfeld enhancement, without appeals to substructure. }
\label{fig:spectra}
\end{figure*}

For $\chi \chi \rightarrow \phi \phi$, we show the resulting spectra for these various $\phi \rightarrow X$ decay modes in Fig: \ref{fig:spectra}. A few points here are in order. We see that essentially all modes give very good fits to the data. Decays to electrons alone ($\phi \rightarrow e^+e^-$, case a) give a very good fit. While $m_\chi = 100$ seems low compared to the highest point, it is only slightly so. This is dominated by the smaller error bars on the lower data points, giving them a much higher statistical weight. It is possible for a 100 GeV WIMP to be within the error bars of the highest data point, but at the expense of a worse overall fit at lower energies. Moreover, we have assumed a flat spectrum for the $\phi \rightarrow e^+e^-$ decay. If $\phi$ is a vector, the injection spectrum can be moderately peaked at high and low energies (low in the middle) \cite{Barger:2008su,Cirelli:2008pk}. Such an effect can improve the fit for the 100 GeV case somewhat, but is essentially irrelevant for higher masses. For $m_\chi \gsim 100$, decays to electrons give comparably good descriptions of the data. 

Decays $\phi \rightarrow \mu^+ \mu^-$ (case b), with the energy partitioned among more particles, requires masses $m_\chi \gsim 400\, \gev$ to provide a good fit to the data. Decays to electrons and muons in a 1:1 ratio (case c) are similar to decays to electrons alone, but with, as expected, a moderately softer spectrum, and a slightly higher boost factor. Decays to charged pions are yet softer, but still provides a good fit to the data, although again, much larger masses and boost factors are required. 

In the highest mass cases, the positron fraction continues to rise above the range where PAMELA can distinguish positrons. The contribution can continue to rise even to the point where the dark matter contribution to the total electronic production is O(1), making deviations in the spectrum of $e^+ + e^-$ very possible at energies O(500 GeV - 1 TeV). Indeed, such an excess may already have been seen at ATIC \cite{ATIC2005} and PPB-BETS \cite{Torii:2008xu}. A comprehensive study combining all data will be performed elsewhere \cite{Cholis:2008wq}.

\vskip -0.25in
\section{Discussion}
Earlier studies \cite{Cholis:2008vb} showed that annihilations $\chi \chi \rightarrow \phi \phi$ could give a good fit to cosmic ray positron excesses seen at HEAT and AMS-01, with the resulting spectra showing sharp rises at higher energies. We have reexamined these predictions in light of the present PAMELA data, and find they provide excellent fits to the data. That decays to dangerous anti-protons or $\pi^0$'s are automatically avoided makes this an extremely compelling explanation for the PAMELA signature arising from WIMPs. Additionally, recent work has further shown that Sommerfeld or other long-distance enhancements can boost the annihilation cross sections for these models, yielding large signals without any conflict with relic abundance constraints.

A WIMP with a mass as low as $\sim$ 100 GeV can give a good fit to the data if $\phi$ decays with a significant component directly to $e^+e^-$. Such a particle requires a very small boost at lower masses, but a large ($B \sim 100$) boost at higher masses. In the context of this scenario, such boosts are expected, however, because of effects from the long distance force, such as the Sommerfeld enhancement.

Similar annihilation rates were shown previously \cite{Cholis:2008vb} to give a good description of the WMAP ``Haze'' \cite{Finkbeiner:2003im,Finkbeiner:2004us,Dobler:2007wv} in the range of $5^\circ-15^\circ$ from the galactic center. As a consequence, a large diffuse ICS signal in the center of the galaxy would be expected to be seen at Fermi/GLAST, particularly for the higher mass particles \cite{Cholis:2008wq}. However, the highest mass scenarios can be constrained by HESS in the event of cuspy profiles \cite{Bertone:2008xr,Bergstrom:2008ag,Mardon:2009rc,Meade:2009rb} from signals in the inner 100 pc, while radio signals can constrain these if such cuspiness continues into the inner 0.1 pc, which may require a flattening of the profile in the inner region \footnote{Some simulations with baryons included point to just such a flattening in the inner 1kpc \cite{RomanoDiaz:2008wz,RomanoDiaz:2009yq}.}.

In summary, a simple modification to the particle physics model - namely, the inclusion of a new light boson - naturally provides a simple explanation for the data. A flattening of the positron fraction in future data from PAMELA may indicate either the mass scale or the decay mode of the $\phi$. Alternatively, the spectrum may continue to rise, in which case a signal may be seen in high energy combined $e^+e^-$ data at PAMELA, or other cosmic ray experiments such as ATIC and PPB-BETS.

\vskip 0.05 in
\noindent {\bf Acknowledgments}
\vskip 0.05in
\noindent We thank N.~Arkani-Hamed, G.~Dobler, J.~Gelfand, D.~Hooper, A.~Pierce and J.~Roberts for helpful discussions. DF is partially supported by NASA LTSA grant NAG5-12972. NW is supported by NSF CAREER grant PHY-0449818, and IC, LG and NW are supported by DOE OJI grant \# DE-FG02-06ER41417.



\bibliography{pamela}
\bibliographystyle{apsrev}

\end{document}